\documentclass[conference]{template/IEEEtran/IEEEtran}

\ifCLASSINFOpdf
\else
\fi
\usepackage[cmex10]{amsmath}

\usepackage{graphicx}
\usepackage{epstopdf}

\usepackage[normalem]{ulem}

\usepackage{color}

\newcommand{\VSP}[0]{\vspace{-0.1cm}}
\begin{document}
%
\title{Multi-Level Modeling of Quotation Families Morphogenesis\\ \vspace{0.5cm} \normalsize Published in the Proceedings of the ASE/IEEE 4th Intl. Conf. on Social Computing "SocialCom 2012"\\Sep. 3-5, 2012, Amsterdam, NL}

\author{
\IEEEauthorblockN{\normalsize Elisa Omodei}
\IEEEauthorblockA{\normalsize LaTTiCe (CNRS, ENS, Paris 3) \\ISC-PIF\\
Email: elisa.omodei@ens.fr}
\and
\IEEEauthorblockN{\normalsize Thierry Poibeau}
\IEEEauthorblockA{\normalsize LaTTiCe (CNRS, ENS, Paris 3)\\
\\
Email: thierry.poibeau@ens.fr}
\and
\IEEEauthorblockN{\normalsize Jean-Philippe Cointet}
\IEEEauthorblockA{\normalsize INRA-SenS\\
INRA, IFRIS, ISC-PIF\\
Email: jean-philippe.cointet@polytechnique.edu}
}


%


\maketitle

\begin{abstract}
This paper investigates cultural dynamics in social media by examining the proliferation and diversification of clearly-cut pieces of content: quoted texts. In line with the pioneering work of Leskovec et al. \cite{Leskovec:2009uf} and Simmons et al. \cite{Simmons:2011wz} on meme dynamics we investigate in deep the transformations that quotations published online undergo during their diffusion. We deliberately put aside the structure of the social network as well as the dynamical patterns pertaining to the diffusion process to focus on the way quotations are changed, how often they are modified and to what extent these changes shape diverse families and sub-families of quotations. Following a biological metaphor, we try to understand in which way \emph{mutations} can transform quotations at different scales and how mutation rates depend on various properties of the quotations.\\
\end{abstract}


%

\section{Introduction}


``Memeticists'', whose forefather Richard Dawkins \cite{Dawkins:2006wm} was the first to coin the term \emph{meme}  in 1976 in analogy to gene, defend the thesis that culture is subjected to evolutionary processes in the same way that living beings are.
This trend of research has suffered strong oppositions especially from ethnographers and anthropologists \cite{sperber1996contagion} criticizing in particular the assumption that culture could be divided into individual objects  called cultural entities \cite{Aunger:2003tn}, {and more generally the lack of empirical validation \cite{edmonds2002three}.
Social media offer at last this opportunity for studying social and  cultural dynamics \emph{in-vivo}}. 
 {In this paper, we claim that tracking the transformations of quotations in the blogosphere is a good way to tackle such empirical endeavour}. If quotations admittedly cannot catch the complex properties found in every cultural traits, their atomic structure is, by nature, an opportunity that should be seized to put the memetic program, at least partially, into practice. 

We are then interested in the diffusion of quotations which builds on the notion of ``intertextuality'', frequently used in political discourse studies. Intertextuality refers to the fact that fragments of discourse are repeated, re-used and progressively modified in different ways. It is thus possible to track the stability or the progressive distortion of an utterance over time. Following Kristeva \cite{Kristeva:1966}, we assume that these distortions are not neutral and reflect the way ideas diffuse in different communities. 

{Diffusion studies have attracted much interest from pioneering sociological approaches \cite{Rogers:1976wl} to more contemporary  studies observing diffusion dynamics in online media. Whatever the nature of the diffusive entity: drugs \cite{Coleman:1957vq},  books recommandations \cite{Leskovec:2007tj}, citations \cite{Leskovec:2007wt}, or URLs \cite{Adar:2004wj}, the process at stake in each of these studies assumes that  objects are perfectly replicated}, their stability being a necessary condition for studying their diffusion. Like the previously mentioned objects, quotations 
can be tracked because whatever their changes they still refer to the same singular external event. But contrarily to them, quotations can undergo transformations, opening the way to the systematic quantitative analysis of regular patterns underlying these changes.

When it comes to characterizing changes that can affect quotations, it may be useful  to follow the biological metaphor provided by memetics. A sequence of genes  can be altered by mutations which may affect only one nucleotide or a large sequence of nucleotides. Small-scale mutations encompass \emph{point mutations} (substitute  a single nucleotide with another), \emph{deletions} and \emph{insertions}. We claim that such an ontology distinguishing between small-scale and large-scale changes is fruitful for addressing quotation transformation dynamics. We then introduce a different typology than the one described by Simmons et al. in their analysis of the same dataset. In \cite{Simmons:2011wz} authors discriminate  between ``reframing'' and ``alteration'' events:
 if a phrase is transformed into a superstring or substring, then reframing takes place, otherwise one should talk about alteration. The relation of inclusion between two quotations  then defines the type of transformation. We prefer to use a different typology directly inspired from the biological mutation process. We will then simply consider on the one hand \emph{micro-mutations} affecting only one word whatever the type of transformation ({\emph{i.e.}} a word can be added, deleted or even replaced by another one) and on the other hand \emph{macro-mutations} affecting more deeply the composition of phrases. Intuitively small-scale and  large-scale mutation events stand for different underlying cognitive processes. Micro-mutations are small changes in the original quotation which can be introduced voluntarily or {simply} by error with no special intention to alter the original meaning of the replicated quote. In the other case, macro-mutations are more probably due to voluntary changes by bloggers or journalists that only want to stress the attention of readers toward a subpart of the original quoted text.

Our goal will then be to describe how micro-scale and macro-scale mutations progressively transform quotations during their diffusion. The first part of the article will be devoted to a very short description of our empirical dataset. An original algorithm for detecting coherent families of quotations is then introduced. {Based on these families, we will then introduce stability and diversity indexes which help us describing the transformation process at different levels (words, quotes and families).} In the last part we will empirically measure mutation rates according to different properties of the quotations and will propose  a  morphogenesis  model for building  realistic families of quotations.
 



\section{Dataset description}
We analyze the MemeTracker corpus which is made of quotations automatically extracted from 90 millions news and blog articles collected over the final three months of the 2008 U.S. Presidential Election and the following three months \cite{Leskovec:2009uf}. 
More precisely, we downloaded the MemeTracker  dataset from the publicly available website \emph{memetracker.org}, that consists of $310\,457$ distinct quotations collected from news and blog articles from August 2008 till the end of January 2009. Each quotation had to be mentioned at least 5 times in order to be included in the corpus.
As we will primarily focus on characterizing how quotes are being transformed, the effects stemming from the underlying social network are out of the scope of this article.   
We then decided to neglect all the hyperlinks between articles and concentrate only on the textual data, i.e. the quotes themselves, and their number of mentions.

%



\section{{Building quotation families}}
 

In order to analyze the MemeTracker corpus, it is necessary to identify  families of quotations, which means grouping together the different quotations in relation with a same ``seed'' quotation (i.e. an original quoted text that can be subsequently re-used, duplicated or modified). 
This analysis is done in three steps: (i) all the quotations are linguistically analyzed and normalized (by lemmatizing the quotes and removing stop words); (ii) similarity between every pair of quotations  is calculated and the quotes whose similarity is above a given threshold are linked so as to obtain a graph of quotations; (iii) a clustering algorithm is applied to detect communities (i.e. cohesive subgraphs in the graph) that will correspond to our families. 
We detail this process in the following subsection, followed by an evaluation of our results and a discussion. 

\subsection{A hybrid linguistic and structural approach}

While the clustering method of Leskovec and his colleagues \cite{Leskovec:2009uf} builds on structural relations between phrases mainly defined according to their potential string inclusion, we tried to design a proximity measure between quoted phrases following more  linguistic hypotheses.

A domain of interest regarding our objective is the paraphrase detection task, which is useful for various natural language applications, including information extraction, automatic summarization and machine translation. Paraphrase detection is highly difficult since it theoretically requires both a semantic and a syntactic analysis of sentences to give valuable results. In practice, most approaches are based on the identification of similar words between couples of sentences, which makes it possible to calculate a similarity value (using a similarity measure like cosine) \cite{mihalcea:2006}. Various refinements can be explored in order to get more accurate results, like trying to calculate word similarity (using for example a resource like Wordnet for English) or trying to identify relations between words. For example, Qiu et al. \cite{qiu:2006} use the Charniak parser to get a syntactic analysis of the sentence and try to map predicate-argument patterns (for example, a verb with its arguments) between sentences, which makes the method more precise. 

In this study, we 
 preferred to design a simple strategy for  building quotation families which features basic text processing techniques and makes use of a refined proximity measure.
First we substituted every word with its lemma using the TreeTagger software \cite{TreeTagger} and eliminated all the stop words.
This step is supposed to conserve only the chore semantic part of each quoted text so that our proximity measure only focuses on the most informative part of each phrase. {Lemmatization allows to unify into one single class simple variations of the same word like singular/plurals, or verbs at different tenses. Stop words, also called ``empty words'', are usually considered as noise when comparing the semantic content of two phrases.}

We  make use of the traditional Levenshtein distance to assess the dissimilarity between two cleaned quotes. But we still need to add some sophistication to take into account word frequency in our measure, considering that rare words are more informative than frequent ones.   
We then computed the tf*idf score \cite{tf.idf}
for every word $w$ of quotation $q$ defined as $\mbox{tf*idf}(w,q,Q) = \mbox{tf}(w,q)* \mbox{idf}(w,Q)$, 
where $\mbox{tf}(w,q)$ is the word $w$ frequency in the quotation $q$ and $\mbox{idf}(w,Q) = \log{ |Q| / \{q \in Q : w \in q\} }$ 
where $|Q|$ is the cardinality of the set $Q$ of all the quotations.
The first term gives more weight to frequent words (in the quotation) and the second one adjusts this value by penalizing words that are too frequent in the dataset 
 since these words are considered to be not discriminative enough.

Then for every couple of quotations $q$ and $q'$ (with $|q| \geq |q'|$) we computed an adjusted Levenshtein distance treating words as tokens and weighting them with their tf*idf scores. Classically, the Levenshtein distance - also called edit-distance - computes the miminum number of additions, deletions or substitutions  necessary to transform an ordered sequence of object into another.
Our weighted Levenshtein distance $\mathcal{L}$  then allows to compare two quotations, defined as two ordered sequences of words following this formula:
\begin{equation*}
 \mbox{$\mathcal{L}$}(q,q') = \frac{ \sum_{\mbox{min edit path}} f(w,q,w',q')(1-\delta(w,w')) }{ \sum_{\mbox{min edit path}} f(w,q,w',q') }
\end{equation*}
where \textquotedblleft min edit path'' is the minimum edit path found by the algorithm to compute the Levenshtein distance, 
$\delta(w,w') = 1$ if $w=w'$, $0$ otherwise, and
\small
\begin{equation*}
f(w,q,w',q') = 
\begin{cases} 
\mbox{max} ( \mbox{tf*idf}(w,q,Q) , \mbox{tf*idf}(w',q',Q) ) 
\\ \;\;\;\; \mbox{if } w = w' \mbox{ or } w \mbox{ substituted } w' \mbox{ or vice-versa} 
\\ \mbox{tf*idf}(w,q,Q) \\ \;\;\;\; \mbox{if } w\mbox{ was inserted or } w' \mbox{ was deleted} 
 \end{cases}
\end{equation*}
\normalsize
{The rationale behind this method is to use a proximity measure based on sequences of words since word order clearly matters (as opposed to bag-of-word approaches where words are considered independently of their order of appearance). But we also give more weight to more discriminating words with their tf*idf score.}

After this pre-processing step, we constructed a similarity network with the set of quotations, in which every quotation is a node. We assign a weighted edge between two quotations $q$ and $q'$ if they have at least two (full) words in common and if their similarity score, calculated as $1-\mathcal{L}(q,q')$,
is greater than $0.35$, a value that we empirically found to be an appropriate threshold. The weight of the edge equals $1-\mathcal{L}(q,q')$.

The final step was to apply an algorithm for community detection in networks in order to identify different quotation families.
For this purpose we chose the Infomap algorithm by Rosvall and Bergstrom \cite{Infomap}, an information theoretic approach algorithm which uses the probability of flow of random walks on a network as a proxy for information flow in the real system and decomposes the graph into communities by compressing a description of the probability flow. Lancichinetti and Fortunato  tested various community detection algorithms \cite{Lancichinetti} and found that Infomap has an excellent performance combined with low computational complexity, which enables to analyze large systems like our dataset.

\subsection{Result evaluation and comparison}

Clustering methods are widely used for natural language processing applications that require grouping different sets of elements. However, evaluating the output of clustering methods remains challenging since gold standards\footnote{In natural language processing, a gold standard is a set of manually annotated data. Most of the time, the data is annotated by several annotators to ensure a reliable result based on a high inter-annotator agreement. } are rarely available and different partitions of the data may often make sense depending on the task and the context. 

As for our experiment, no gold standard was available but it is possible to use the result of the MemeTracker project  experiment as a comparison point. We chose to rely on a manual evaluation of a relevant sample of clusters randomly selected from those produced by our method and those produced by the MemeTracker method.
We randomly selected 30 of our families, and for each family we also selected  every MemeTracker family that had at least one quotation in common with the original family. Then for each family we made two lists: the first one containing all the quotations in the family, and the second one containing all the quotations which belonged to the corresponding selected families of the MemeTracker project. Then we did the opposite with 30 families of the MemeTracker project.
The size of the initial clusters used for evaluation varies from 3 to 150 text snippets/sentences. 

The list was then assessed by two judges, who were told that every first list represents 
{a subset of closely related quotations} and to mark any quotation that they thought should not belong to the family. Then they had to look at the second list and mark if any of the quotations  should be added to the family.

Before detailing the result of this evaluation, it is necessary to quickly examine some methodological issues. First, a number of text snippets were not real quotations but titles (``high school musical''), short expressions with no clear meaning out of context (``a little bit'') or foreign words (``la vida no vale nada'') between brackets. The corresponding clusters were excluded from the evaluation\footnote{Note that our families will be filtered in a second phase in order to delete these kinds of pathologic families.}. Second, the identification of the main information expressed in a set of snippets is a difficult task, especially given the variation in length of the different snippets. The instruction given to the evaluators was to first have a look at the whole set of snippets before determining the prominent information, which seems to have worked pretty well. Lastly, the instruction was to tag as equivalent snippets that were reporting the same main information even if some secondary information was missing. It was possible to tag a snippet as uncertain. 

Despite the minimal set of instructions given to our evaluators, we obtained interesting and reliable results. We compared the evaluation produced by two annotators and obtained a high inter-annotator agreement (Cohen's kappa is $0.69$, which is surprisingly high given the relative subjectivity of the task and the scarcity of instructions provided to the evaluators). 

We obtained the following results for precision and relative recall (the  recall is  relative since we performed an evaluation based on a comparison between two methods and not with respect to a gold standard), where precision is calculated as the fraction of quotations considered relevant in the first list, recall as the number of relevant quotations in the first list over the same number plus the number of relevant quotations found in the second list. 

\begin{center}
\begin{tabular}{|l|r|r|r|}
  \hline
  Clustering Method & Precision & Relative Recall & F-measure \\
  \hline
  {Ours} & .58 & .90 & .70 \\
  MemeTracker & .47 & .78 & .58 \\
  \hline
\end{tabular}
\end{center}
 
Our method outperforms Leskovec et al.'s approach both in precision and relative recall. This increase in precision is probably due to the linguistic preprocessing step that makes the whole process more precise (our analysis is focused on content words that are themselves weighted according to their discriminative power). The increase in relative recall is probably due to the fact that the MemeTracker clusters contain also a lot of snippets that are very small fractions of the seed quotation and were thus considered by the judges not to carry  enough information to be unambiguously part of the family.

We also measured the statistical significance between the two F-measure values through the SIGF V2 test \cite{sigf06}, which implements an assumption-free randomization framework. It allows to assess whether the difference in performance between two sets of predictions is significant. 
We found a p-value of $0.049$, which means that the F-scores are different with a $5\%$ significance level.

\subsection{Family filtering}
Since the dataset contains many quotations that are either not in English either too short to convey a real unit of meaning, we first decided to filter it by considering only quotations containing at least 5 words in English\footnote{For this purpose we used the \textit{guess-language} software available at http://code.google.com/p/guess-language/.}.

\subsection{Family description}

\begin{figure}
\begin{center}
\includegraphics[scale=0.34,angle=270]{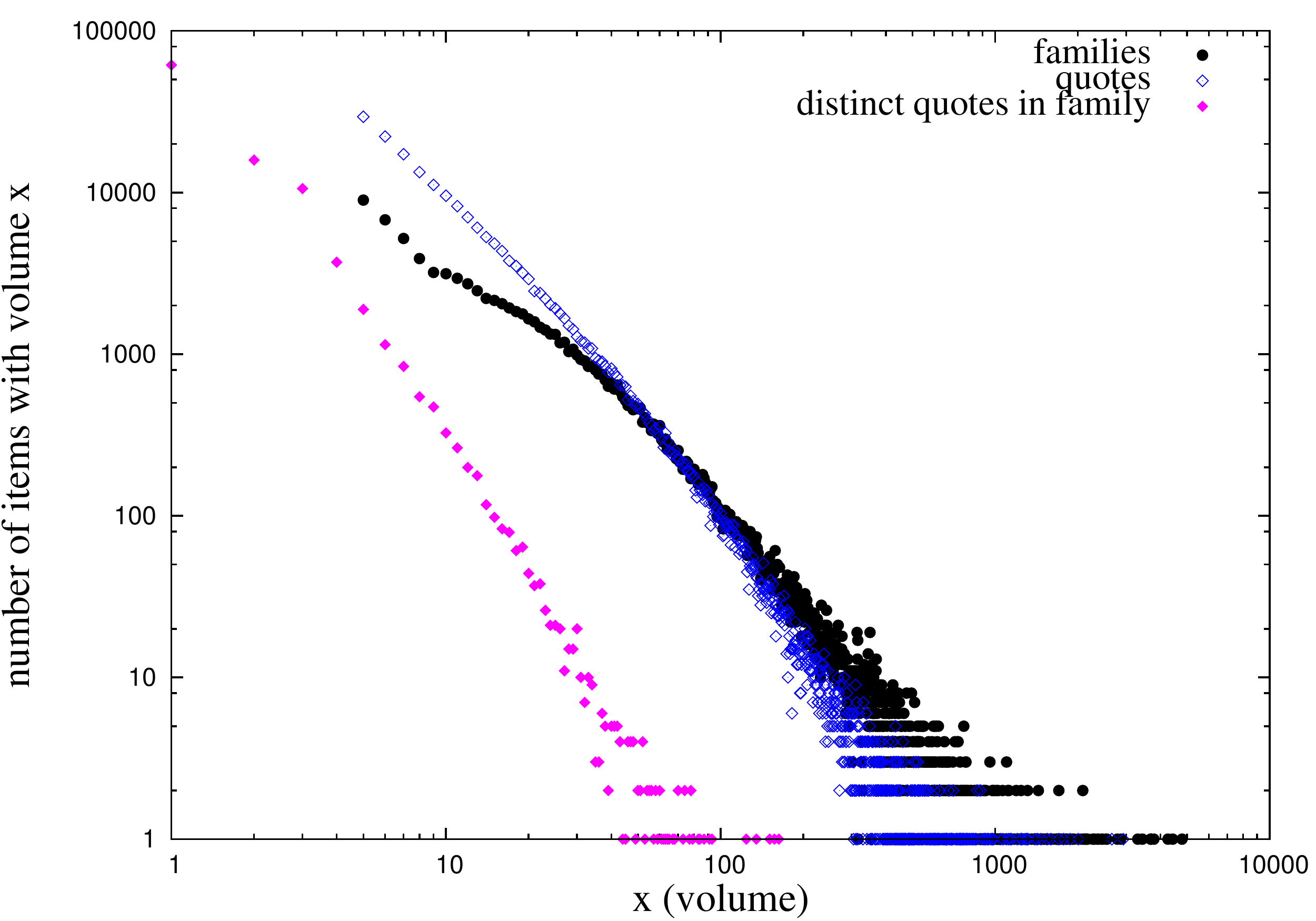} 
\end{center}
\caption{Distribution of family sizes defined as the total number of mentions of the quotations in the family (black dots), distribution of the number of distinct quotations in a family (pink full diamonds) and distribution of the number of mentions per quotation (blue diamonds).}
\label{dsitribution-size}
\end{figure}

We plotted the distribution of family sizes defined as the total number {of quotations  mentioned in a family}, the distribution {of the number of distinct versions found in each family} and the distribution of the number of mentions per quotation (Fig.~\ref{dsitribution-size}). 
The shapes of all the three distributions can be approximated as a power-law and are comparable to distributions found in  \cite{Leskovec:2009uf} {although the families were  defined differently}. 

\section{Multi-level transformation analysis}

\begin{figure}
\begin{center}
\includegraphics[scale=1.24]{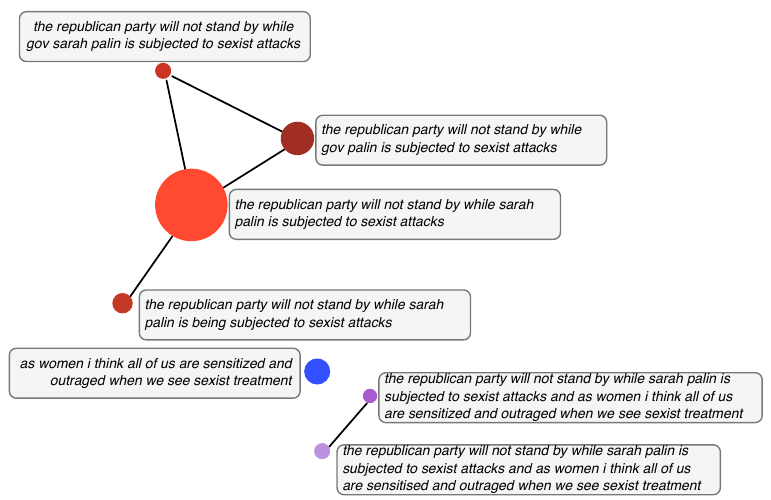} 
\end{center}
\caption{Empirical family detected by our algorithm and its associated edit-distance graph. Quotes are connected when their edit distance is at most $1$. Sub-families (colored in red, blue and purple in this example) are made of connected components of the edit graph. Node sizes scale with the observed total number of mentions of the corresponding quote.}
\label{example}
\end{figure}

Before going into further details, let's first have a look at an actual family identified as a family gathering 7 different quotes from the MemeTracker dataset. On Sep 3, 2008, Carly Fiorina, former boss of computer-making giant Hewlett-Packard, told a press conference: ``\emph{The Republican party will not stand by while Sarah Palin is subjected to sexist attacks ... and as women, I think all of us are sensitized and outraged when we see sexist treatment}''. The quote is genuinely replicated $16$ times in the dataset {but it can also be found in 6 alternative forms, the most frequent of them being mentioned $56$ times in a much shorter form than the original one: ``\emph{the republican party will not stand by while sarah palin is subjected to sexist attack}''.}

Observing the diversity of quotes within the family, it appears that 3 main sub-families clearly emerge gathering either the first, the second or both parts of the original quote. Within each sub-family, we also observe small variations between different versions of the quotations due to different spellings of the same word (\emph{sensitised}/\emph{sensitized}), words trimmed or added (\emph{sarah palin}/\emph{gov palin}/\emph{gov sarah palin}, \emph{is being subjected}/\emph{is subjected}).

As we wish to take into account the {diversity of possible changes}, we will distinguish in the following between \emph{small-scale transformations} observed when a single word is changed, added or removed, and \emph{large-scale transformations} occurring when the quote is significantly trimmed into a smaller version.

\subsection{Definitions}

{Before providing a more formal definition of these transformations}, we first define the edit-distance graph as the network connecting quotes whose edit-distance is no larger than 1. {Edit distance is  defined as a Levenshtein distance between two quotes considering words as single characters.  The edit distance between two quotes is then the minimum number of edits needed to transform one quote into the other, with the allowable edit operations being insertion, deletion, or substitution of a single term.} The maximum number of allowed edits is fixed to one\footnote{We have chosen to consider only the edit-distance graph connecting quotes at the smallest possible edit-distance \emph{i.e.} $1$. However, we have checked that our results {are qualitatively unchanged when considering} larger transformations (edit distance at most $2$ or $3$).}. Please note that in this case and in all the following analysis no preliminary linguistic processing is applied to quotes before computing their edit-distance. Contrarily to the strategy we adopted for detecting families, we naturally {do not wish to alter quotations at this step  as we are interested in identifying every possible transformation.}

For every family, we then build the edit-distance graph $G$ connecting its quotations, and extract its connected components. These connected components define the sub-families, \emph{i.e.} sets of quotes which can only be differentiated by small-scale changes. Applied to our former example, edit distance graph indeed allows us to exhibit three different sub-families including various micro-level variants (see Fig. \ref{example}).
The edit distance graph is not only useful to define sub-families since it will also be used to identify - given a target quote $q$ -  which quotes are found in its immediate neighbourhood $\mathcal{V}_q$\footnote{$\mathcal{V}_q$ will define the set of quotes whose edit-distance with $q$ is less than or equal to $1$ and that belong to $q$'s family. Note that $q \in \mathcal{V}_q$.} (i.e. quotes that are directly competing for the attention of bloggers or journalists).
{We then introduce three different measures that will help us to assess the transformation dynamics at different levels.}
\paragraph{Term level}
we are first interested in the relative stability of terms\footnote{In this study we define terms as  simple words.} found in a quotation. 
Given a quote $q$ and a term $t\in q$, we define the stability of this term in this quote $s(t|q)$ as the proportion of quotes in  $\mathcal{V}_q$ that share the same term\footnote{More formally the  stability of $t$ in a given quote $q$ is defined as: $s(t|q)= \frac{\sum_{q' \in \mathcal{V}_q, t \in q'}w_{q'}}{\sum_{q' \in \mathcal{V}_q}w_{q'}}$, where $w_q$ stands for the total number of mentions of $q$.}.
The global stability of a term $t$ is then defined as the weighted average of term stabilities computed for every quote it belongs to, that is to say:
$s(t)=\sum_{q,t\in q}w_q s(t|q)$, where $w_q$ stands for the total number of mentions of $q$.

\paragraph{Quote level}
at the quote level, we simply define the stability of a quote $q$  as the proportion between the number of mentions of $q$  and the total number of mentions of quotes in its immediate neighborhood:
$S(q)= \frac{w_{q}}{\displaystyle\sum_{q' \in \mathcal{V}_q}w_{q'}}$. 
{At this stage, we voluntarily focus on  micro changes, excluding large-scale transformations that may occur when  copying  only a subpart of a quotation for example.}
\paragraph{Family  level}

we also wish to appraise how much a family or a sub-family composition is heterogeneous. We compute the entropy of the distribution of number of mentions for every quote in the family / sub-family: 
$H_F=-\sum_{q\in F}p_q log(p_q)$ where $p_q$ holds for the proportion of {mentions of} quotes $q$ in its family / sub-family: $p_q=\frac{w_q}{\sum_{q \in F} w_q}$.

\subsection{Term level}
The observation of the list of the most unstable words exhibits some well known patterns, the first one being the slight orthographic variations that exist for certain words in English. We thus observe the following alternations: ``defence/defense'', ``programme/program'' and, among many others, ``behaviour/behavior''. Other variations include words with a dash (``cease-fire/ceasefire''), abbreviations (``gov/governor'') and foreign words (``al-qaeda/al-qaida''). Lastly, slang words are frequently omitted, which makes them more subject to variation than ordinary words (for example, ``fuck'' is in the top 20 most instable words in the corpus; it can be either suppressed or replaced by a simple ``f''  {in indirect quotes}). 

Besides these qualitative observations, more systematic patterns emerged when analyzing term stability according to different properties.
We also wish to describe which features can provide quotes or terms higher-fidelity replication rates. Put differently, we are asking which properties at the term or quote level may systematically account for a higher or lower mutation rates?  Formally, the stability of any feature attached to a term/quote is defined as the weighted average of every terms/quotes sharing this property\footnote{For example for assessing the stability of terms with a given total  frequency in the corpus $n$  we will compute: {$s(f=n)=\frac{\sum_{t,f(t)=n}\sum_{q, t \in q} w_q s(t|q)}{\sum_{t,f(t)=n}\sum_{q, t \in q} w_q}$}}. 

\begin{figure}
\begin{center}
\includegraphics[scale=0.34,angle=270]{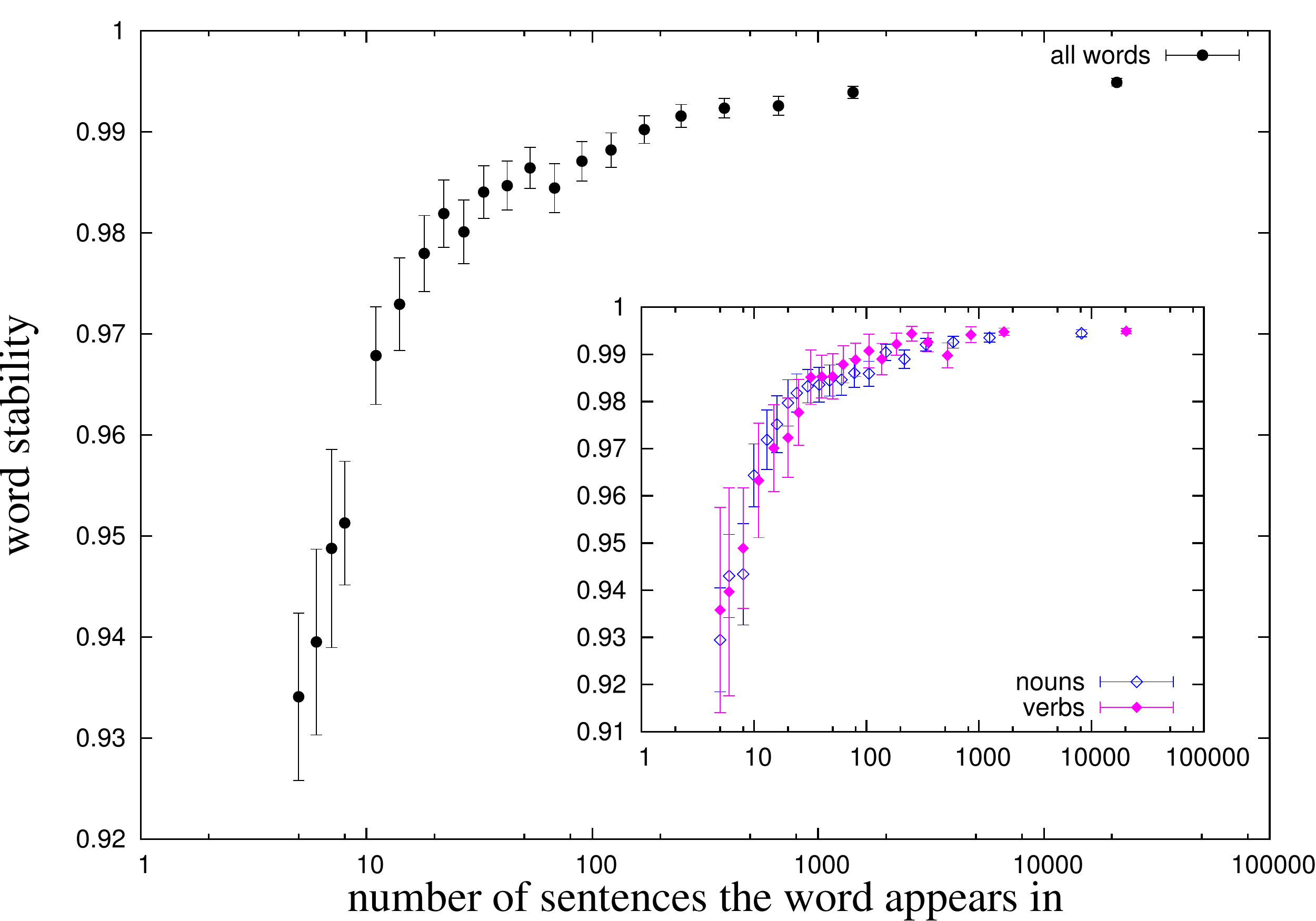} 
\end{center}
\caption{Stability in function of term frequency. The figure is obtained by creating 20 equally populated quantiles and averaging the stability values corresponding to each quantile. Error bars stand for confidence interval ($5\%$). \emph{Inset}: stability of verbs and nouns only.}
\label{wordstab-wordfreq}
\label{gridbrain}
\end{figure}
Figure~\ref{wordstab-wordfreq} shows that word frequency significantly affects term stability. 
More frequent terms are significantly more likely to be stable.  More precisely terms with more than $100$ occurrences in our quote corpus have a stability higher than $99\%$, this value reaching up to $99.5\%$  for the most frequent terms. The rarer the term the more dramatically its stability falls. We checked that this pattern is still present after removing stop-words from our set of terms. The same pattern can be observed even when selecting only certain grammatical types of terms, suggesting that frequency of terms plays a central role when it comes to memory issues or when one has to decide whether a given term should undergo a change. This observation may seem counter-intuitive if one considers that less frequent terms may convey more specific meaning. Yet rare terms may also be more prone to change as they may be misspelled or simply more difficult to spell precisely because of their scarcity. The same kind of observation has actually been made in studies examining the long-term evolution of language.  For example, in  \cite{Lieberman:2007vq}, authors showed that the regularization rate of irregular English verbs was rapidly decreasing with their frequency, indicating that low frequency irregulars verbs are subject to more errors, leading to their ``rapid'' regularization. 
{ Besides, specificity is not synonymous with stability. A recent study analyzing the same dataset used  \emph{Wordnet}\footnote{Wordnet is a lexical database of English featuring synonymous relations between words (http://wordnet.princeton.edu/).}  to rank terms according to their genericity. They showed that the more specific the terms the more likely they are to be replaced, especially by more generic ones \cite{lerique2012}. This ``natural preference'' for more generic terms may  also account for the shape of the curve we observe as it is probable that more specific terms occur less frequently than more polysemic ones.}

\begin{figure}
\begin{center}
\includegraphics[scale=0.45]{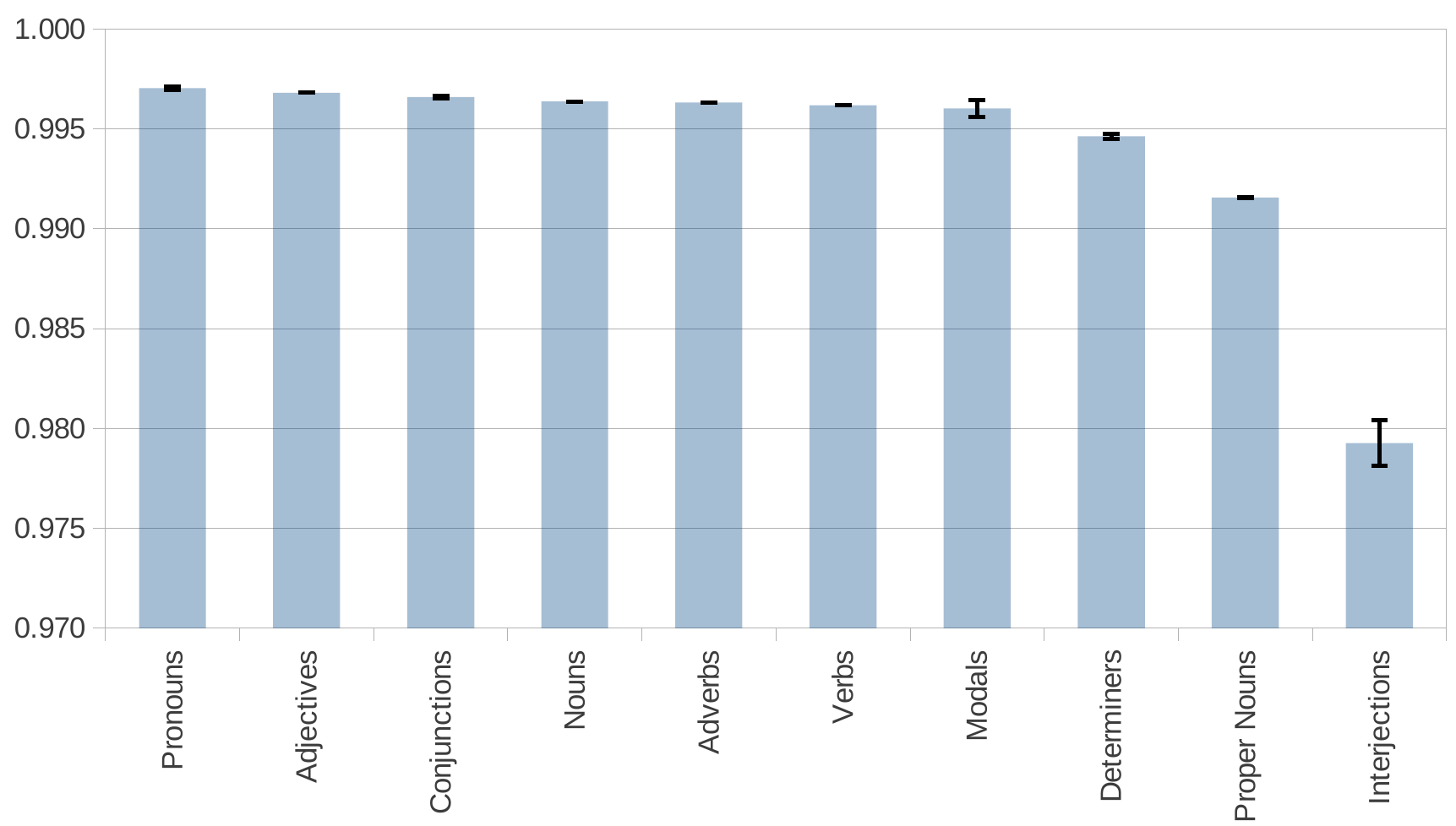}
\end{center}
\caption{Stability values for different grammatical types.}
\label{grammatical}
\end{figure}

We also computed the average stability of terms according to their grammatical type (see Fig.~\ref{grammatical}). Results are as expected.  
Most common grammatical types approximately have the same level of stability. Yet we note that besides interjections which we could expect to feature lower stability, proper noun tend to be more than twice more unstable than average. Indeed proper nouns may produce more mutations as they can be misspelled or undergo more transformations as illustrated in our example in Fig.~\ref{example}.

\subsection{Quote level}

\begin{figure}
\begin{center}
\includegraphics[scale=0.34,angle=270]{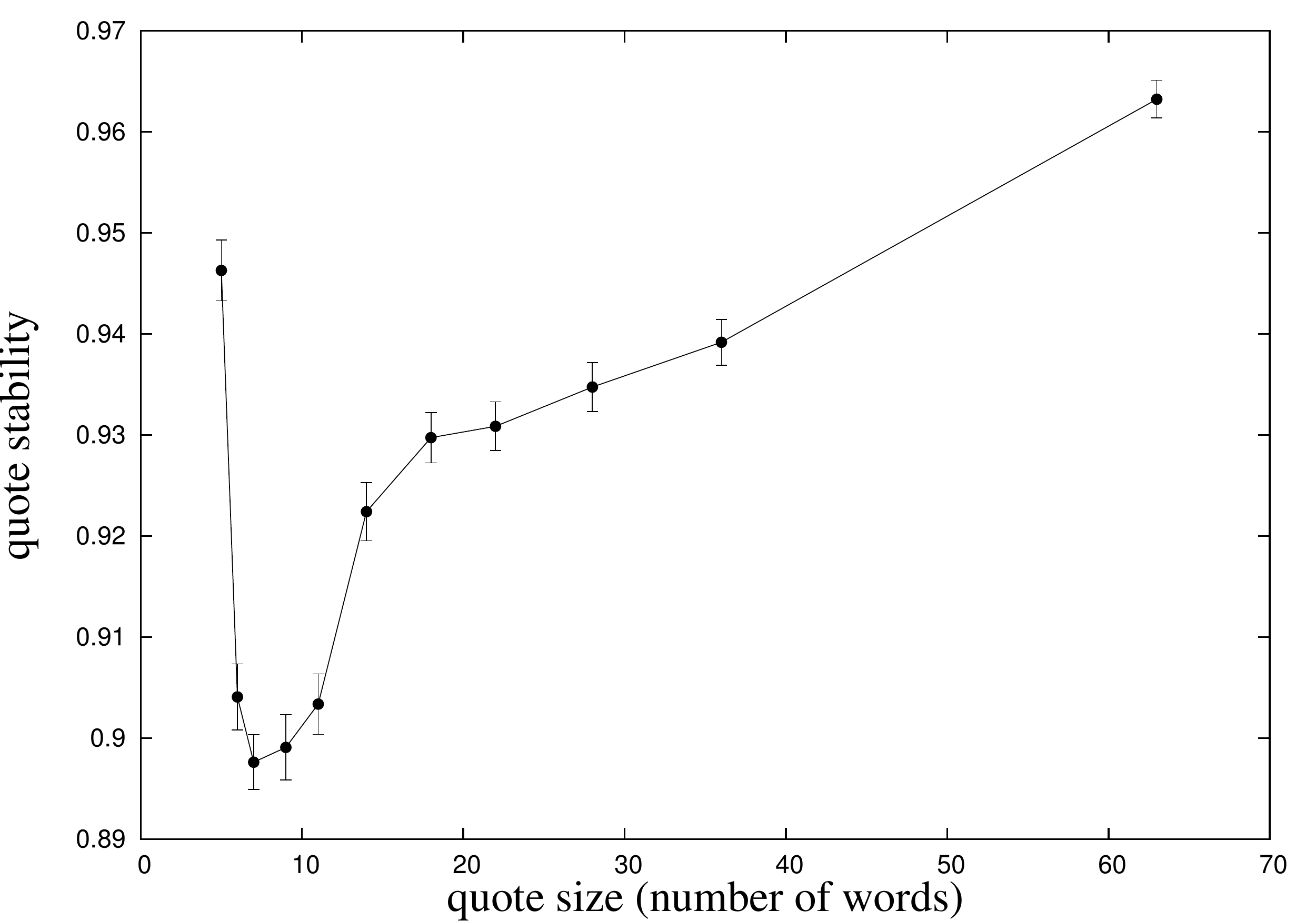} 
\end{center}
\caption{Quote stability in function of quote length (number of words). The figure is obtained by creating 10 equally populated quantiles and averaging the stability values corresponding to each quantile. Error bars stand for confidence interval ($5\%$).}
\label{quotesize}
\end{figure}

We computed quote stability and plotted stability against quote length, i.e the number of words that it contains (see Fig. ~\ref{quotesize}).
Quote stability is minimal for a certain length (around 8 words). Smaller quotes are less keen to change, while longer quotes also feature higher stability. On such a digital medium, two processes may be in competition when it comes to editing quotes: \textit{i}) a blogger may read/hear a quote in a newspaper, Twitter, or more broadly catch it from any media and try to replicate it by memory or \textit{ii}) he can simply copy \& paste the quote from a digital source. The first copy process is probably more used for quotes that are not too long (a 10-15 words long quote seems already quite difficult to memorize), and is also more keen to introduce variations than a pure copy \& paste process (note that variations depend on the length of the quote: very short quotes are easier to memorize and are thus quite stable; longer quotes are more unstable, 8 words long quotes being the most unstable ones). On the other hand, it seems plausible that longer quotes have greater chances to have been replicated from an existing source (for quotes over 8 words long, copy \& paste is probably the preferred option, hence the increase in stability for longer quotes). 
This competition between low and high-fidelity replication along with the uneven probability to introduce errors according to the size of the quote may explain the particular shape of the curve. 

\begin{figure}
\begin{center}
\includegraphics[scale=0.34,angle=270]{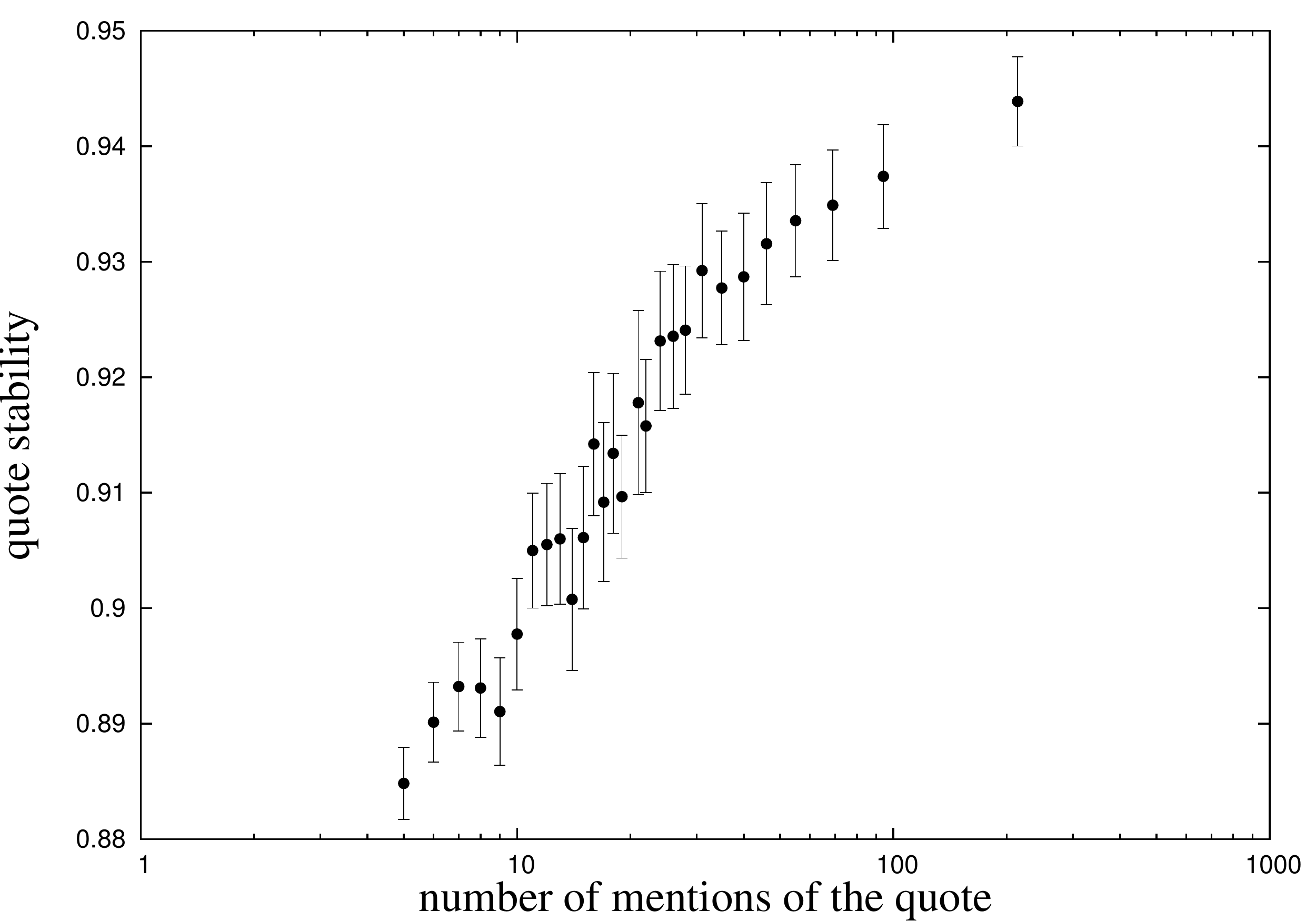} 
\end{center}
\caption{Quote stability in function of their number of mentions. The figure is obtained by creating 30 equally populated quantiles and averaging the stability values corresponding to each quantile. Error bars stand for confidence interval ($5\%$).}
\label{freqquote-stab}
\end{figure}

We also investigated whether a quote stability is affected or not by its total frequency. In Figure~\ref{freqquote-stab} we plotted the weighted average stability of quotes according to their frequency and observe that the curve increases logarithmically. 
Two explanations may account for the higher stability of high frequency quotes. People trying to replicate them may produce less errors simply because they are inherently better replicators (this property also explaining their popularity). Another cause of their stability may simply be that they are being copied - because of their spread - significantly more often than their alternative forms, increasing in the long-run the disparity between their frequency and alternative forms which are less and less likely to get copied.
The two processes may also be taking place at the same time: popular quotes tend to be naturally copied more frequently, while their number may decrease the chances to introduce mistakes in the copying process as it would seem more unlikely for someone to alter a quote she/he has already read several times. 

\subsection{Family / sub-family level}

\begin{figure}
\begin{center}
\includegraphics[scale=0.34,angle=270]{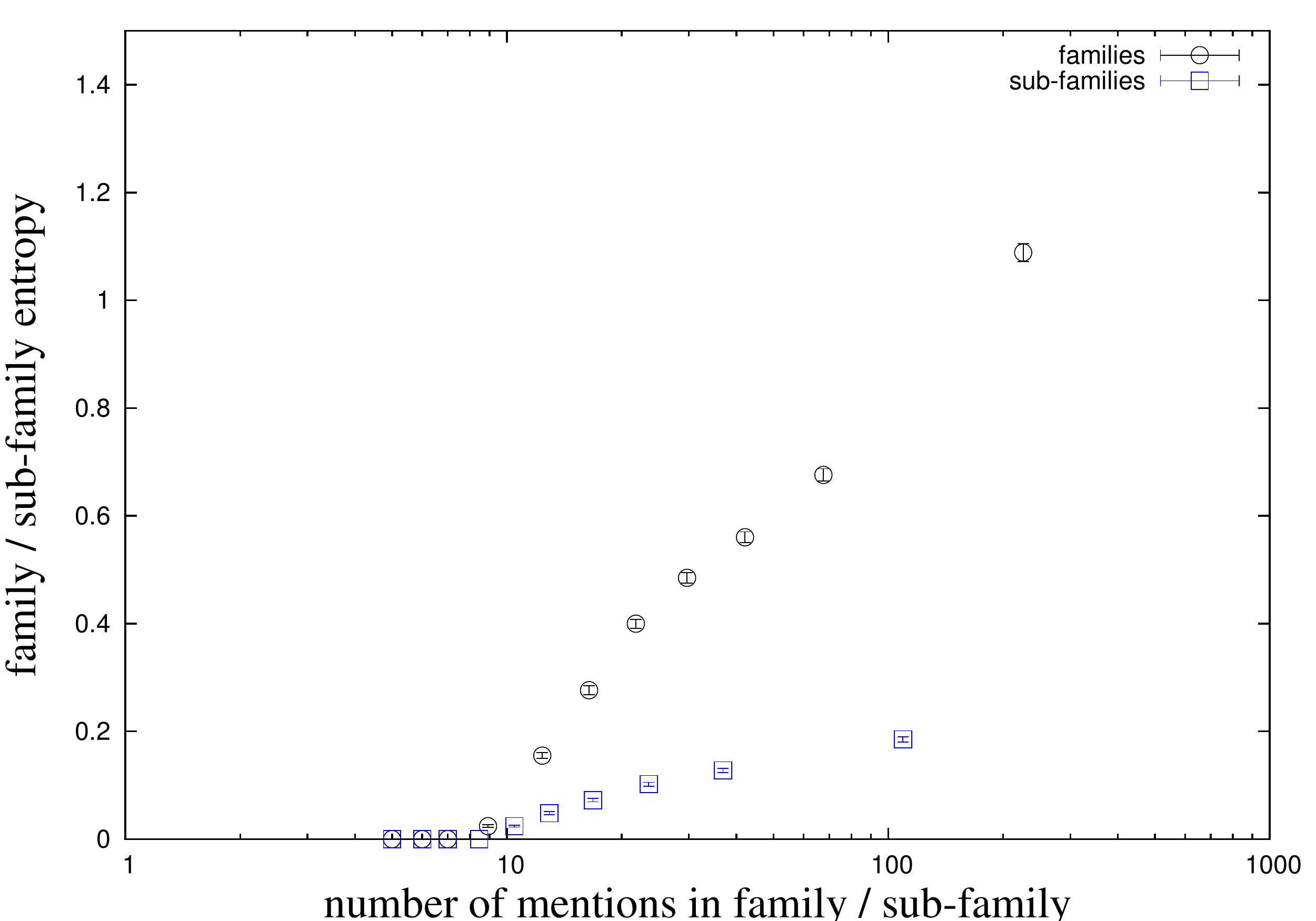} 
\end{center}
\caption{Family and sub-family entropy in function of the total number of mentions of their quotes. The figure is obtained by creating 10 equally populated quantiles and averaging the entropy values corresponding to each quantile. Error bars stand for confidence interval ($5\%$).} 
\label{entropy}
\end{figure}

To better understand how families are composed and more precisely their inner diversity we also measured their entropy. We both measured the entropy of sub-families and of families as a whole. We recall that families group togther every quote related to an original quote whatever the scale of transformations it may have undergone, while sub-families gather quotes from the family that can be connected through micro-level changes.
The Shannon entropy, which was originally applied to letters \cite{Shannon:vd}, is classically used as a diversity index. Applied to quotes, the entropy measures the diversity of quotes composing a family or a sub-family. 
Entropy at the sub-family and family levels exhibits very different patterns. Figures~\ref{entropy} shows how entropy correlates with the family / sub-family size.
{We  observe that the value of the entropy for the families with a certain number of mentions is always much higher than the value of the entropy of the sub-families of the dataset gathering the same number of mentions. Put differently sub-families exhibit less diversity than families, suggesting that - at the micro-level - the competition among different versions of the same phrase eventually leads to a situation in which there is one version that is predominant regarding the number of mentions with respect to other versions, whereas - at the macro-level - there is more heterogeneity due to the coexistence of different relatively independent sub-families.}

\section{Modeling Quotation Family Generation}

We now  propose a model of quotation family morphogenesis that accounts for their composition in terms of sub-family size distribution, and regarding their diversity at both levels. To design a realistic model, we still need to precisely define how quotations are being changed when the family is growing.  
In this section we will then have a closer look at the temporal evolution of families, examining how many and which type of mutations are introduced during the process.

\subsection{Mutation rates}
We have shown that two types of mutations could occur when copying a quote. When the quotation is not perfectly replicated, one may  observe a macro-level mutation - only a subpart of the original quotation is selected - or a micro-level mutation - small changes affecting only one word in the quotation. 
 In the first case, we will consider that a new sub-family  is produced, in the second case that the sub-family the original quote belongs to is simply enriched by a new version. We make the assumption that the chances that a mutation produces a version that had been already published before is negligible. 

As we assume that every new version is triggered by a mutation event, we can assess mutation rates {\emph{a posteriori}} at both levels by {comparing the number of different versions in a given subset and the total number of mentions these different versions received. More precisely, the average mutation rate can be computed at both micro / macro level  as the ratio between the number of micro / macro-mutation events (number of versions  in the sub-family minus one / number of sub-families in the family minus one), and the total number of copying steps (mentions in the sub-family / family minus one).}

But accessing average mutation rates is not enough to realistically reconstruct family and sub-family morphogenesis: we also need to take into account the relevant properties affecting mutation rates. {In the previous section, we showed  that quote stability is modified according to their  length and their popularity, which suggests that mutation rates could strongly differ according to those two properties}. Besides, those properties may critically depend on the diffusion dynamics. That is the reason why we should    
{\emph{dynamically} assess  the rate at which new versions are being produced in the empirical process according to those different conditions.}

{We then define the following strategy to dynamically measure mutation rates. Each family is considered as a growing set progressively populating the various sub-families with new quotes. Each time a new quotation 
is produced, we record whether it is a perfect copy of a previously mentioned quotation, or a new version that had never been observed before. In the latter case we also record whether the original quotation is enriching an existing sub-family or creating a completely new sub-family. We compile those events according to the original state of the family and sub-family\footnote{{We make the hypothesis that a new quotation enriching a pre-existing sub-family was necessarily copied from one member of this sub-family 
}}, \emph{i.e.} we enumerate   micro-changes, macro-changes and perfect copying events according to the average sub-family size 
and average quotation length. 
From there on, it is straightforward to define the micro / macro-mutation rate according to a given average length or a given total number of mentions as the proportion of replication events producing a micro / macro change. We will call the so computed micro and macro mutation rates $\rho_\mu$ and $\rho_M$ respectively.

\begin{figure}
\begin{center}
\includegraphics[scale=0.34,angle=270]{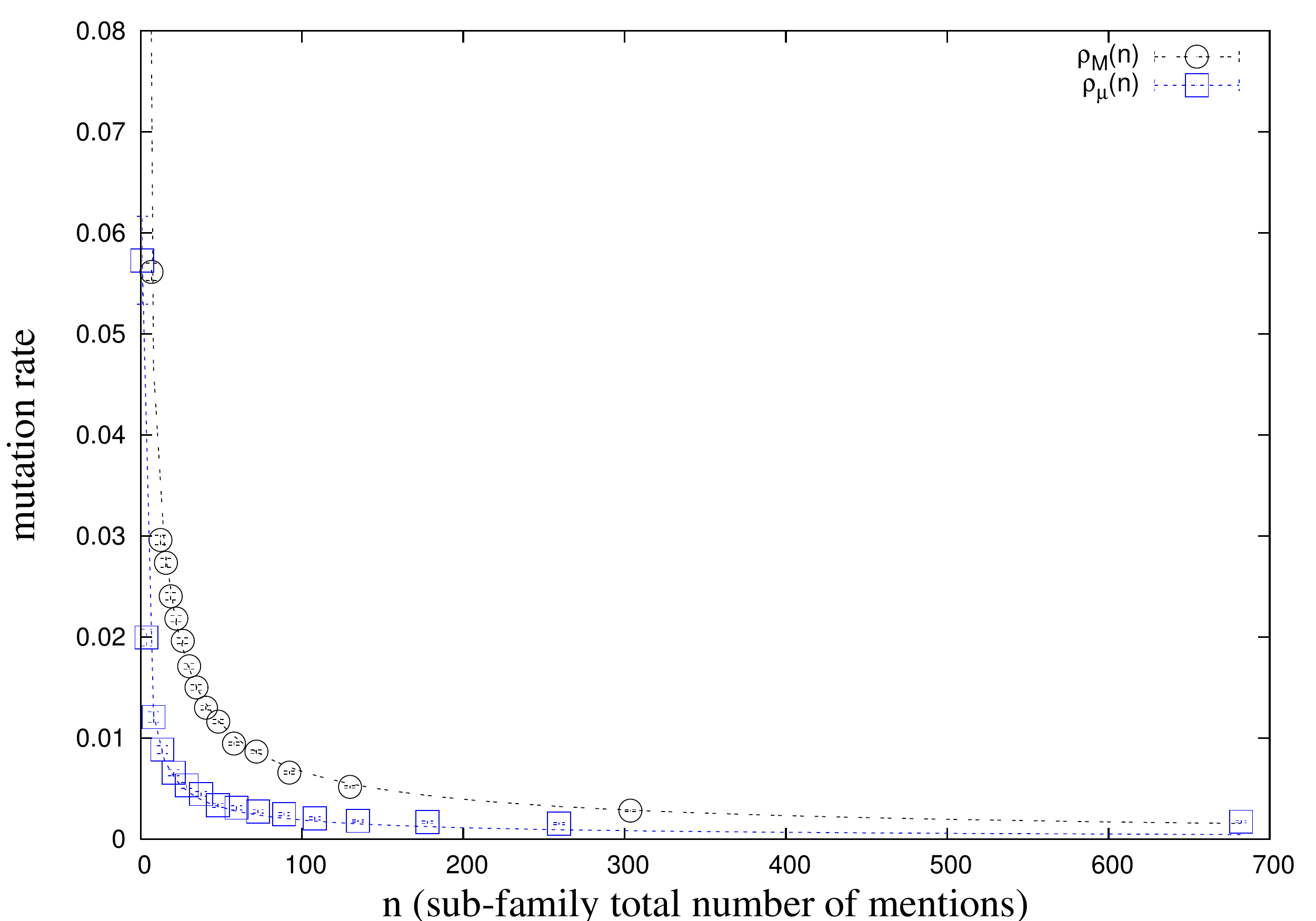}
\end{center}
\caption{Micro and Macro mutation rates according to the sub-family total number of mentions, along with their fitted model. $\rho_M(n) = 0.225n^{-0.763}$ and $\rho_\mu(n) = 0.057n^{-0.739}$. The figure is obtained by creating 15 equally populated quantiles and averaging the mutation rate values corresponding to each quantile. Error bars stand for confidence interval ($5\%$).}
\label{muocc}
\end{figure}
From figure~\ref{freqquote-stab} we can suspect that  the number of mentions is crucial for determining the precise mutation rate of a quotation. Therefore we plotted (Fig.~\ref{muocc}) the micro (and respectively  macro) mutation rates according to {the total number of mentions observed in the  sub-family}. As expected we find that both mutation rates decrease with the number of mentions 
  (see figure caption for further details about the fitting functions used)}.
 This behaviour confirms our hypothesis that more popular quotations are less keen to changes. Very popular quotations may be so ubiquitous in the  environment that the probability to introduce micro-mutations by error is lowered (many copies can recall the agents how the correct version should be spelled) or that ``successful'' quotations have such {high} ``fitness'' that any further refinements  is unnecessary.

\begin{figure}
\begin{center}
\includegraphics[scale=.34,angle=270]{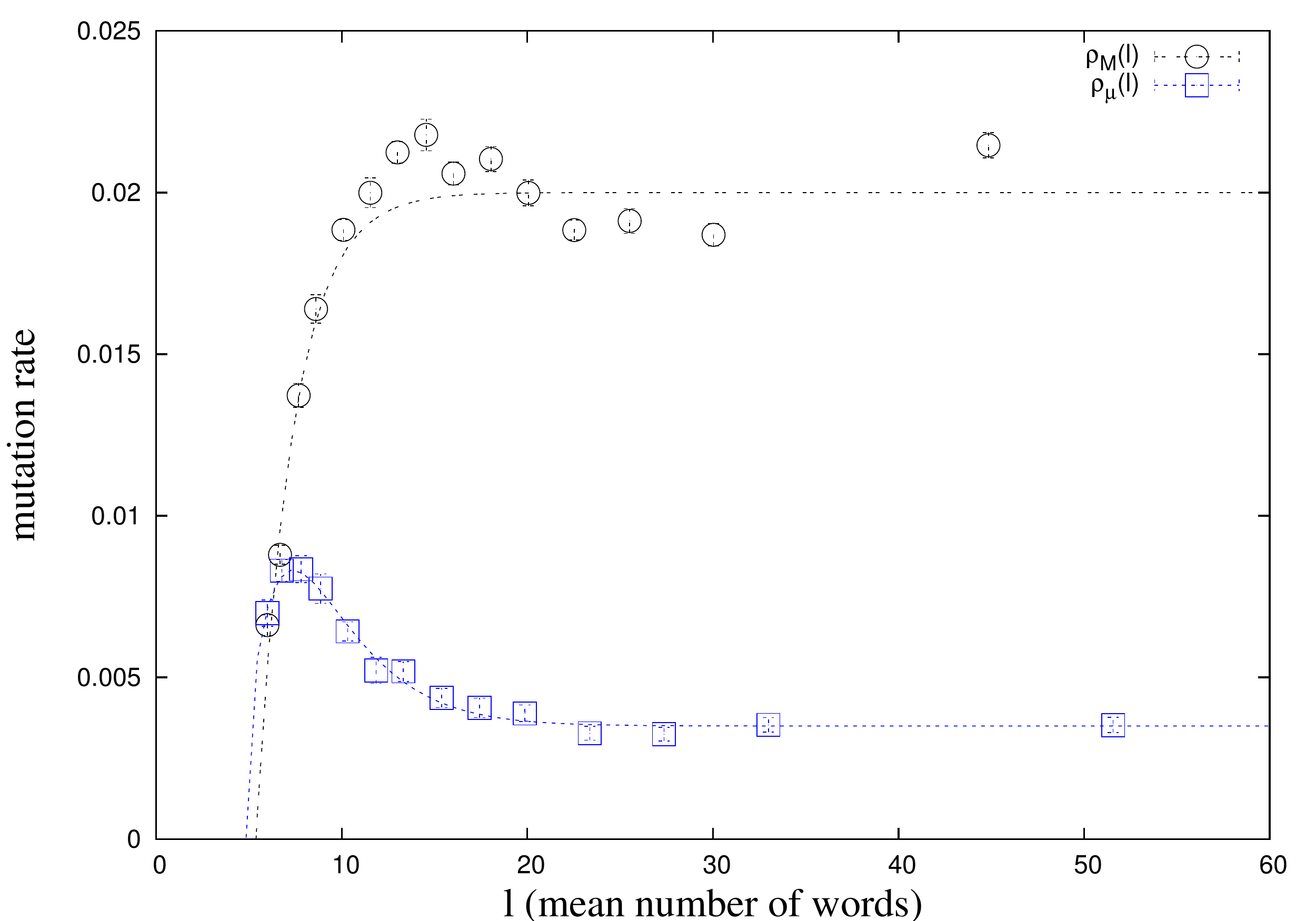} 
\end{center}
\caption{Micro and Macro mutation rates according to quotation lengths, along with their fitted model. $\rho_M(l) = 0.020-0.292 \exp {(-0.499l)}$ 
and $\rho_\mu(l) = 0.004+0.046(l-5) \exp {(-0.423l)}$. 
The figure is obtained by creating 15 equally populated quantiles and averaging the mutation rate values corresponding to each quantile. Error bars stand for confidence interval ($5\%$).}
\label{mutationlen}
\end{figure}

Figure~\ref{quotesize} showed that quotation stability is sensibly modified according to their length $l$. We plotted in Fig.~\ref{mutationlen} the mutation rates {according to the average length of the family / sub-family quotations}. 
We observe that the macro-mutation rate is growing with quotation length.  While small quotations (less than $5$ words) can naturally not undergo any macro mutation given the definition of our family categorization, the macro-mutation rate reaches a threshold for quotations over $20$ words. In our model, we use an exponential function to express the dependence of the macro mutation rate with quotation size (see Fig.~\ref{mutationlen} caption for further details). {A plausible explanation is that a quotation can hardly be trimmed before reaching a certain critical length}. Above $20$ words the quotation is certainly made of different phrases or sentences that individually carry some autonomous meaning even when separated from their original environment.  

The correlation between $l$ and the micro mutation rate seems more complex. A peak of the micro mutation rate is reached for mid-size quotations around 8 words. After the maximal value is reached, we observe an exponential decrease until the curve plateau at the minimal mutation rate.
As already hypothesized when commenting the shape of quotation stability with length, micro-change dynamics seems to be driven by two processes acting in different directions regarding {the number of words}. First, it seems clear that mistakes introduced during the copying process are possible only when the quotation is not simply copy \& paste. It seems reasonable to postulate that the probability for a quotation to be  retrieved from memory rather than copy \& paste  is exponentially  decreasing with quotation length. 
If the quotation was retrieved from memory then chances are that some mistakes will be introduced. If we refer to classical works in psychology \cite{Miller:1956tr}, human brain can hold up to a certain number of objects or chunks in memory. This so-called ``magic number'' below which short term memory is almost perfectly accurate is precisely around $5$ for words.   
This is the reason why we chose to fit the correlation between micro mutation rate and length with a more complex equation made of the product of two probabilities: the probability that the quoted phrase is not replicated by a copy \& paste event (which is exponentially decreasing with $l$) and the probability that an error is introduced by chance (which is assumed to be linear for quotations larger than the magical number $5$). This product models the probability that the quotation is replicated with an error. We also add a baseline in the fitting function to account for the constant probability that a blogger or a journalist voluntarily introduces a single change to the quotation (see Fig.~\ref{mutationlen} caption for further details).


\subsection{Model Design}
\begin{figure}
\begin{center}
\includegraphics[scale=.8]{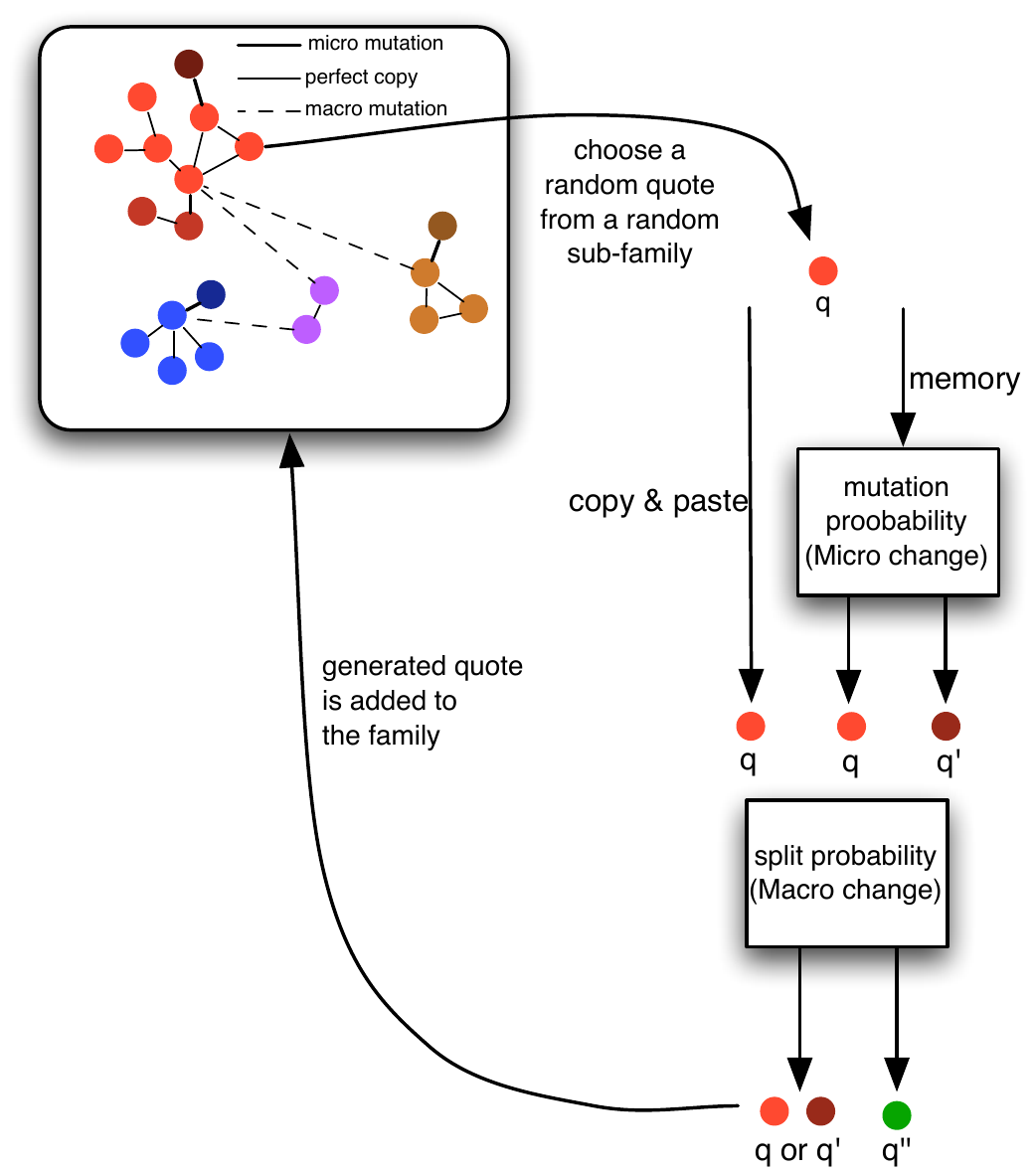} 
\end{center}
\caption{{Schematic} representation of the model dynamics.}
\label{model}
\end{figure}

We now propose a generative process producing families of quotations. Our aim is to find a realistic agent-based process that accounts for the  distribution of the size of subfamilies as well as for the shape of the increase of diversity at the family and sub-family level in time. 
We rely on a classical Polya urn principle like in \cite{Simmons:2011wz}. We assume that each sub-family is centered around a specific kernel of meaning and is then characterized with its own autonomous dynamical process. This is the reason why we randomly select a sub-family from which a quotation {is picked for replication. }
{ The quotation may then undergo a} mutation according to its length $l$ and its number of  mentions $n$.  

{More formally,  a  family $F$ is initialized with a quotation $q$ of a given length $l$ ($F=\{q\}$). This  first quotation is also assigned a sub-family $f$. The simulation then iterates over every time step as follows (see Figure~\ref{model} for an illustration): }
\begin{enumerate}

\item One randomly chooses a sub-family $f$ of $F$ and then a quotation $q \in f$ with probability proportional to the number of mentions of $q$
\item With a probability given by the combination of the two micro mutations rates $\rho_\mu(l)$ and $\rho_\mu(n)$\footnote{
{Precisely the global probability to observe a mutation given the total number of mentions and average length is given by $\rho = \frac{\rho(l)\rho(n)}{<\rho>}$.{ As one can not simply infer which was the original quote from which a new quote was replicated, we computed the stylized behaviour linking mutation rates with $l$ and $n$ considering average lengths and estimated cumulated number of mentions in the sub-family. We are now making the hypothesis that  mutation rates can be  directly computed based on the quotation length and number of mentions. Lengths being homogeneously distributed, the approximation seems reasonable.  Since the distribution of mentions is heterogeneous and given that our process preferentially selects the most cited quote, it is very likely that the number of mentions of a random quote is well approximated by the total number of mentions in the sub-family.}}} the quotation undergoes a micro-change, resulting in a new quotation $q'$ that differs from the original one by only one edit (deletion, insertion or substitution). If above this probability the quote is not modified,
\item Then, if long enough, the quotation can also be trimmed into a smaller quotation with a probability given by the the macro mutation rates which depend on the quotation length $l$ and  number of mentions $n$ according to the fitted values of $\rho_M(l)$ and $\rho_M(n)$\footnotemark[\value{footnote}].
{ If so,  a new shorter} quotation $q''$ is created.
\item {If the quotation did not undergo any mutation on steps $2$ and $3$,}  {it} is perfectly replicated (copy\&paste) and a new quotation $q$ is produced.
\item The possibly mutated ($q'$ or $q''$) or unchanged ($q$) version of the original quotation is added to the family. 
\end{enumerate}
{The process is repeated from step $1$ until  the family is considered complete, \emph{i.e.} when it has received the total number of quotations we observed in the empirical distribution.}




\subsection{Model Results}
\begin{figure}
\begin{center}
\includegraphics[scale=0.34,angle=270]{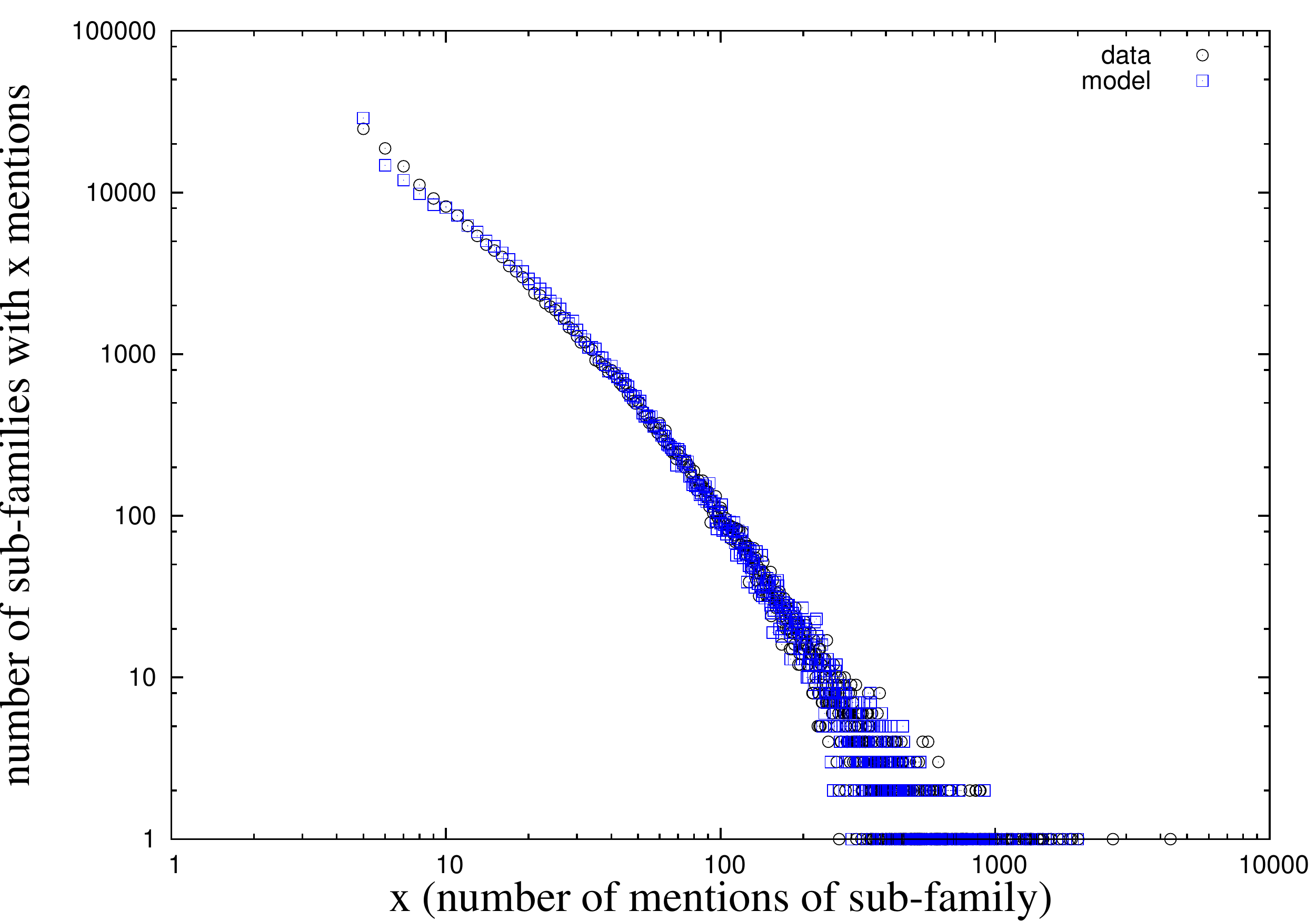} 
\end{center}
\caption{Comparison of the distribution of sub-family sizes in the empirical data (black dots) and produced by the model (blue diamonds).
}
\label{subfamdistr}
\end{figure}
\begin{figure}
\begin{center}
\includegraphics[scale=0.34,angle=270]{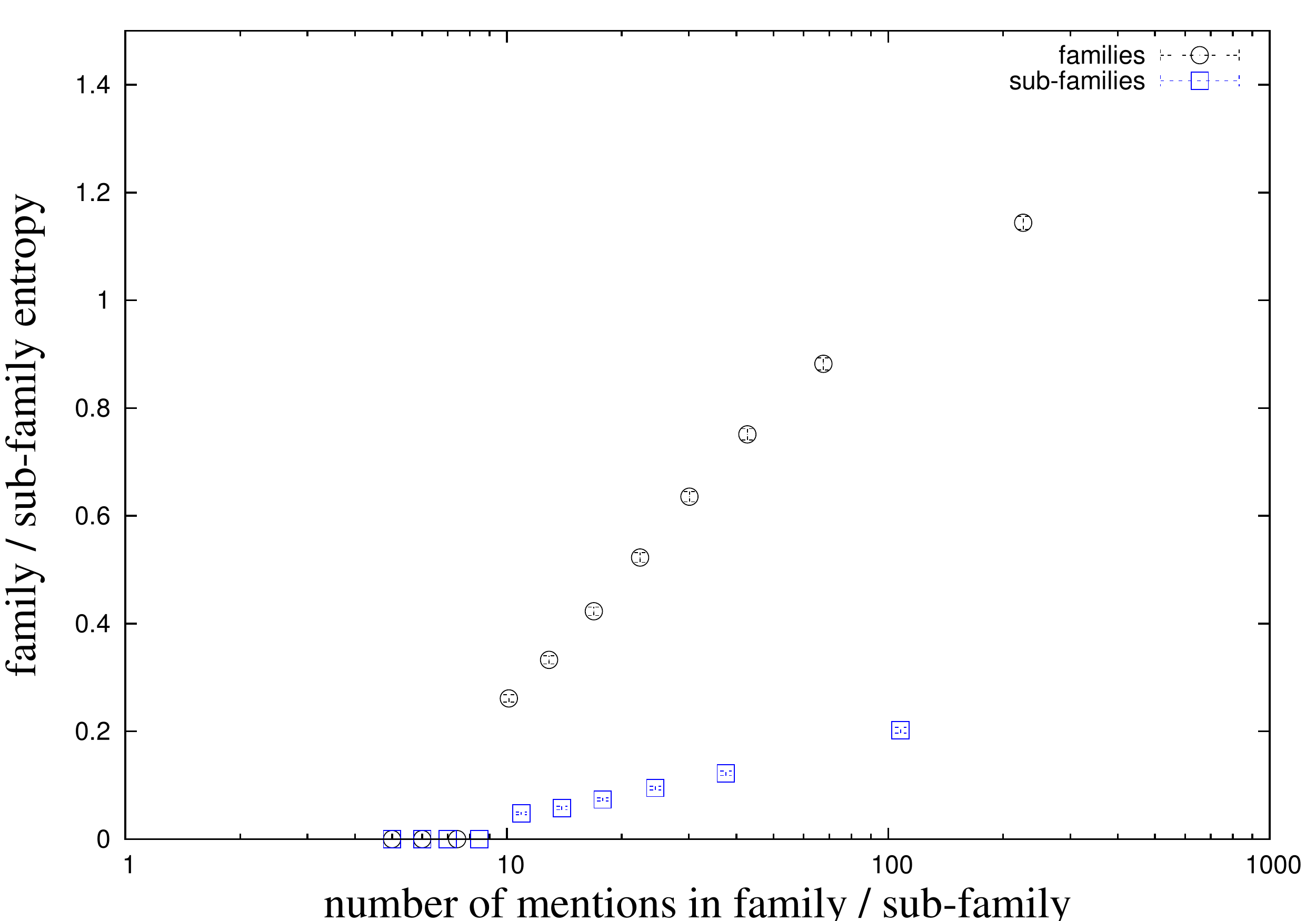} 

\end{center}
\VSP
\caption{Family and sub-family entropy produced by the model in function of the total number of mentions of their quotations. The figure is obtained by creating {10 equally populated quantiles}. Error bars stand for confidence interval ($5\%$).}
\VSP
\label{entropy_model}
\end{figure}

{Our simulation almost produced the same number of distinct versions and sub-families than in our original dataset {(less than $1\%$ error)}.} {We also observe that} the proposed model accounts for the size distribution of sub-families and for the diversity of families and sub-families. Figure~\ref{subfamdistr} shows a very good {fit} between the empirical and the simulated sub-family size distributions, suggesting that our model succeeds in reproducing the observed feature that families are composed of sub-families of more similar quotation versions. Moreover, Figure~\ref{entropy_model} shows that our model is also able to reproduce the difference that we observed in the values of entropy between families and sub-families. The model creates quite homogeneous sub-families that put together into families account for their larger diversity.


\section{Conclusions}

In this paper we introduced a new algorithm for quotation clustering as a first  step towards the analysis of quotation family structure and transformations. We showed how these families can be characterized at a meso level by retrieving the corresponding sub-families, i.e. the connected components of a graph of edit distances of at most one edit. This multi-level analysis allowed us to find new interesting results, such as the difference in the entropy level at the two scales, suggesting that the {strong} competition among very similar quotations leads to a more homogeneous situation with respect to the {co-existence  of} the different sub-families. Moreover, we presented a model that attempts to describe the morphogenesis of these families of quotations and accounts for their composition in terms of sub-family size distribution and for the {difference of diversity measured at both levels}. The model relies on the analysis of quotation stability, through which we showed that quotations undergo a different number of mutations according to their length and {their number of mentions}, concluding that quotations that are already very popular have less chances to be transformed.
In future work we would like to take into account the underlying social network in order to enrich the analysis of the {driving forces determining}   quotations transformation. Moreover, our model {could be significantly more realistic  with a finer description of temporal patterns pertaining to quotation diffusion.}\\
%
%



\section*{Acknowledgment}
The authors would like to thank Tommaso Brotto and Zorana Ratkovic for serving as judges for the clustering method evaluation. We also warmly thank the reviewers as well as S\'ebastien Lerique, Camille Roth, Andre\"\i ~ Mogoutov, {Benjamin Fagard, Isabelle Tellier} and Jonathan Platkiewicz for fruitful comments. Elisa Omodei's work is supported by a PhD grant from the R\'egion Ile-de-France.
This work has also been partially supported by the French National Agency of Research (ANR) through the grant “Algopol” (ANR-12-CORD-0018).




\begin{thebibliography}{10}
\providecommand{\url}[1]{#1}
\csname url@samestyle\endcsname
\providecommand{\newblock}{\relax}
\providecommand{\bibinfo}[2]{#2}
\providecommand{\BIBentrySTDinterwordspacing}{\spaceskip=0pt\relax}
\providecommand{\BIBentryALTinterwordstretchfactor}{4}
\providecommand{\BIBentryALTinterwordspacing}{\spaceskip=\fontdimen2\font plus
\BIBentryALTinterwordstretchfactor\fontdimen3\font minus
  \fontdimen4\font\relax}
\providecommand{\BIBforeignlanguage}[2]{{%
\expandafter\ifx\csname l@#1\endcsname\relax
\typeout{** WARNING: IEEEtran.bst: No hyphenation pattern has been}%
\typeout{** loaded for the language `#1'. Using the pattern for}%
\typeout{** the default language instead.}%
\else
\language=\csname l@#1\endcsname
\fi
#2}}
\providecommand{\BIBdecl}{\relax}
\BIBdecl

\bibitem{Leskovec:2009uf}
J.~Leskovec, L.~Backstrom, and J.~M. Kleinberg, ``{Meme-tracking and the
  Dynamics of the News Cycle},'' \emph{Proceedings of The Fifteenth ACM SIGKDD
  International Conference on Knowledge Discovery and Data Mining (KDD-09)},
  2009.

\bibitem{Simmons:2011wz}
M.~Simmons and L.~Adamic, ``{Memes Online: Extracted, Subtracted, Injected, and
  Recollected},'' \emph{ICWSM 2011}, 2011.

\bibitem{Dawkins:2006wm}
R.~Dawkins, \emph{{The selfish gene}}.\hskip 1em plus 0.5em minus 0.4em\relax
  Oxford University Press, USA, 2006.

\bibitem{sperber1996contagion}
D.~Sperber, \emph{La contagion des id{\'e}es}.\hskip 1em plus 0.5em minus
  0.4em\relax Jacob, 1996.

\bibitem{Aunger:2003tn}
R.~Aunger, ``{Cultural transmission and diffusion},'' \emph{Encyclopedia of
  cognitive science}, 2003.

\bibitem{edmonds2002three}
B.~Edmonds, ``Three challenges for the survival of memetics,'' \emph{Journal of
  Memetics-Evolutionary models of information transmission}, vol.~6, no.~2, pp.
  45--50, 2002.

\bibitem{Kristeva:1966}
J.~Kristeva, ``Word, dialogue and novel,'' in \emph{The Kristeva Reader},
  T.~Moi, Ed.\hskip 1em plus 0.5em minus 0.4em\relax Oxford: Blackwell, 1966.

\bibitem{Rogers:1976wl}
E.~M. Rogers, ``{New Product Adoption and Diffusion},'' \emph{The Journal of
  Consumer Research}, 1976.

\bibitem{Coleman:1957vq}
J.~S. Coleman, E.~Katz, and H.~Menzel, ``{The Diffusion of an Innovation Among
  Physicians},'' \emph{Sociometry}, 1957.

\bibitem{Leskovec:2007tj}
J.~Leskovec, L.~A. Adamic, and B.~A. Huberman, ``{The dynamics of viral
  marketing},'' \emph{portal.acm.org}, 2007.

\bibitem{Leskovec:2007wt}
J.~Leskovec, M.~McGlohon, C.~Faloutsos, N.~Glance, and M.~Hurst, ``{Cascading
  behavior in large blog graphs},'' \emph{SIAM International Conference on Data
  Mining (SDM 2007)}, 2007.

\bibitem{Adar:2004wj}
E.~Adar, L.~Zhang, L.~A. Adamic, and R.~Lukose, ``{Implicit structure and the
  dynamics of blogspace},'' \emph{Workshop on the Weblogging Ecosystem, 13th
  International World Wide Web Conference}, 2004.

\bibitem{mihalcea:2006}
R.~Mihalcea, C.~Corley, and C.~Strapparava, ``Corpus-based and knowledge-based
  measures of text semantic similarity,'' in \emph{Proceedings of the American
  Association for Artificial Intelligence (AAAI 2006)}, Boston, 2006.

\bibitem{qiu:2006}
L.~Qiu, M.-Y. Kan, and T.-S. Chua, ``Paraphrase recognition via dissimilarity
  significance classification,'' in \emph{Proceedings of the 2006 Conference on
  Empirical Methods in Natural Language Processing}, Sydney, Australia, 2006.

\bibitem{TreeTagger}
H.~Schmid, ``Probabilistic part-of-speech tagging using decision trees,'' in
  \emph{Proceedings of International Conference on New Methods in Language
  Processing, Manchester, UK}, 1994.

\bibitem{tf.idf}
G.~Salton and M.~McGill, \emph{Introduction to modern information
  retrieval}.\hskip 1em plus 0.5em minus 0.4em\relax McGraw-Hill, 1983.

\bibitem{Infomap}
M.~Rosvall and C.~T. Bergstrom, ``Maps of random walks on complex networks
  reveal community structure,'' \emph{PNAS}, 2008.

\bibitem{Lancichinetti}
A.~Lancichinetti and S.~Fortunato, ``Community detection algorithms: A
  comparative analysis,'' \emph{Phys. Rev. E}, 2009.

\bibitem{sigf06}
S.~Pad\'o, \emph{User's guide to \texttt{sigf}: Significance testing by
  approximate randomisation}, 2006.

\bibitem{Lieberman:2007vq}
E.~Lieberman, J.~Michel, J.~Jackson, T.~Tang, and M.~Nowak, ``{Quantifying the
  evolutionary dynamics of language},'' \emph{Nature}, vol. 449, no. 7163, pp.
  713--716, 2007.

\bibitem{lerique2012}
S.~Lerique and C.~Roth, ``How do brains copy and paste? the semantic drift of
  quotes in blogspace.'' \emph{Forthcoming}, 2012.

\bibitem{Shannon:vd}
C.~Shannon, ``{A Mathematical Theory of Communication (pt. 1), 27 Bell Sys.
  Tech. J. 379 (1948); CE Shannon}.''

\bibitem{Miller:1956tr}
G.~Miller, ``{The magical number seven, plus or minus two: Some limits on our
  capacity for processing information.}'' \emph{Psychological review}, vol.
  101, no.~2, p. 343, 1956.

\end{thebibliography}

%
%
%

\end{document}